\documentclass[twocolumn,showpacs,amsmath,amssymb]{revtex4}

\usepackage{graphicx}
\usepackage{dcolumn}
\usepackage{bm}
\newcommand{\etal}{\emph{et al.}}
\newcommand{\be}{\begin{equation}}
\newcommand{\ee}{\end{equation}}
\newcommand{\bfig}{\begin{figure}}
\newcommand{\efig}{\end{figure}}
\newcommand{\incl}{\includegraphics}

\begin{document}      

\title{Anomalous thermopower and Nernst effect in $\rm CeCoIn_5$: entropy-current loss in
precursor state}

\author{Y. Onose$^1$, Lu Li$^1$, C. Petrovic$^2$ and N. P. Ong$^1$
}
\affiliation{
$^1$Department of Physics, Princeton University, Princeton, NJ 08544, USA\\
$^2$Department of Physics, Brookhaven National Laboratory, Upton, N.Y. 11973, USA
}
\date{\today}      
\begin{abstract}
The heavy-electron superconductor CeCoIn$_5$ exhibits a puzzling precursor state
above its superconducting critical temperature at $T_c$ = 2.3 K.
The thermopower and Nernst signal are anomalous.  Below 15 K, the entropy current 
of the electrons undergoes a steep decrease reaching $\sim$0 at $T_c$.  Concurrently, the off-diagonal thermoelectric current $\alpha_{xy}$ is enhanced.  The delicate 
sensitivity of the zero-entropy state to field implies phase coherence over large
distances.  The prominent anomalies in the thermoelectric current contrast with 
the relatively weak effects in the resistivity and magnetization.
\end{abstract}

\maketitle                   
Several novel, interesting phenomena have been discovered in the anisotropic, heavy-electron 
superconductor CeCoIn$_{5}$ which has a critical transition temperature $T_c$ = 2.3 K~\cite{petrovic}.  
Evidence for $d$-wave pairing symmetry are seen in 
a number of experiments~\cite{movshovich,izawa,ormeno,aoki}. 
The Fulde-Ferrell-Larkin-Ovchinikov (FFLO) state has been shown to exist below a 
temperature $T\sim$ 0.3 K~\cite{FFLO}. In a magnetic field $\bf H || c$ (the $c$-axis), 
a crossover from non-Fermi liquid to Fermi liquid behavior has been reported~\cite{FL,onose}. 
Magnetization results~\cite{mag} reveal that a weak, unusual magnetic order begins to appear
$\sim$20 K above $T_c$.  This precursor state is mysterious and
often associated with a ``hidden'' order parameter.
Recently, Bel \etal~\cite{bel} reported that a giant Nernst signal appears
at $\sim$25 K.  They reported that the sign of the Nernst signal is
opposite to that expected from the vortex-Nernst signal,
which has been intensively studied in cuprates~\cite{xu,wang01,wang06}.

Using thermopower, Nernst-effect and other transport measurements 
on high-purity CeCoIn$_5$, we have determined
the full thermoelectric (Peltier) conductivity tensor {\boldmath $\alpha$}.
In the precursor state at $H$ = 0, a pronounced reduction 
of the thermopower reveals a remarkably steep loss of carrier entropy current.  
A weak field suppresses this low-entropy state.  After subtraction of the 
thermal-Hall contribution to the Nernst signal and correcting its sign, 
we show that the enhanced Nernst signal correlates with the thermopower
anomalies in the precursor state.

An applied temperature gradient $-\nabla T$ drives 
the charge current density $\bf J = ${\boldmath $\alpha$}$\cdot(-\nabla T)$.  
Because the total current density in the sample is zero,
an internal electric field $\bf E$ arises to drive a counter 
current $\bf J' = ${\boldmath $\sigma$}$\bf\cdot E$, where {\boldmath $\sigma$}
is the conductivity tensor.
The $x$ and $y$ components of $\bf E$ are observed, respectively, 
as the thermopower $S = E_x/|\nabla T|$ and the Nernst signal $e_N = E_y/|\nabla T|$
(we choose axes with $-\nabla T||\hat{\bf x}$ and $\bf H||\hat{z}$).
The thermopower, which involves balancing 2 counter charge currents $||\bf \hat{x}$,
is just the ratio of 2 transport quantities, i.e. $S = \alpha/\sigma$,
with $\alpha \equiv\alpha_{xx}$ and $\sigma \equiv\sigma_{xx}$ (we drop subscripts on
diagonal quantities). However, the Nernst signal $e_N$ is more involved.

In the simplest situation, the transverse gradient $-\partial_y T$ is negligible
(isothermal case).  Along $\bf\hat{y}$, we have the 2 off-diagonal 
currents $\alpha_{yx}(-\partial_x T)$ and $\sigma_{yx}E_x$, which must be
cancelled by $\sigma E_y$.  In terms of the resistivity 
tensor $\mbox{\boldmath $\rho$} = \mbox{\boldmath $\sigma$}^{-1}$, we have~\cite{wang01}
\be
e_N = \rho\alpha_{xy} + \rho_{xy}\alpha.
\label{en}
\ee
The Nernst signal $e_N$ senses the sum of the off-diagonal Peltier current
and the ordinary Hall current multiplied by $S$.

\bfig[h]			
\incl[width=9.5cm]{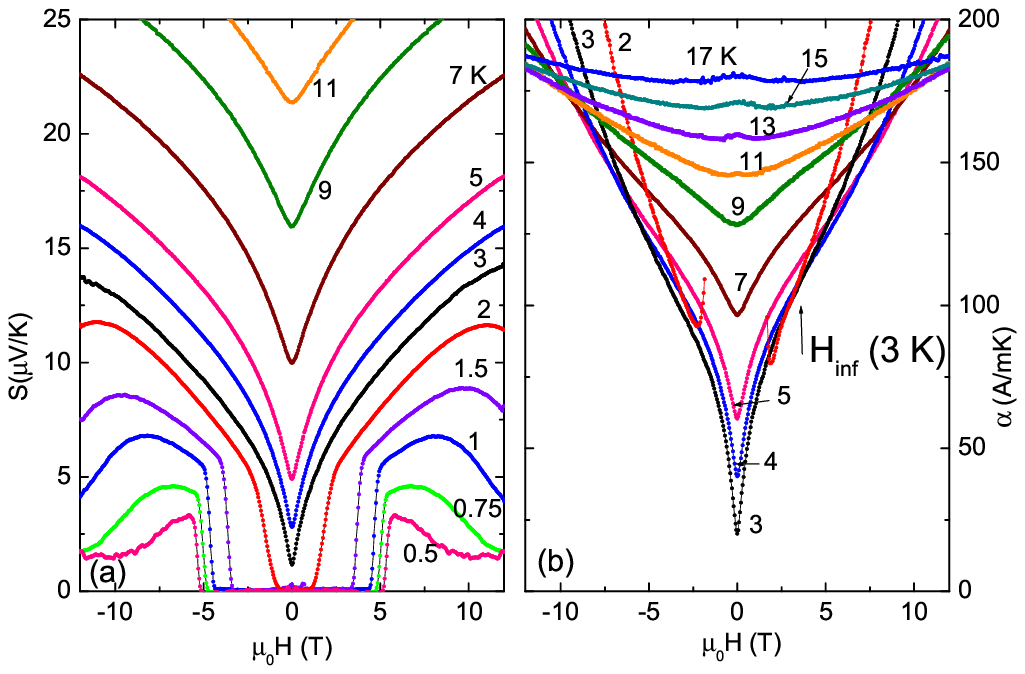}
\caption{\label{SH}  (Panel a) The field dependence of the thermopower $S(H)$ of CeCoIn$_5$ 
at selected $T$ ($\bf H||c$). The zero-field anomaly deepens rapidly as $T$ fall towards
$T_c$ = 2.3 K. Below $T_c$, $S(H)$ rises nearly vertically at
$H_{c2}$, but continues to increase to a broad maximum (at $\sim$8 T at 1 K). Beyond the
peak $S$ decreases to a plateau value. Panel (b) shows curves of 
$\alpha$ vs. $H$ from 2 to 17 K.  As $T\rightarrow T_c$, $\alpha$ decreases
(increases) if $|H|$ is smaller (larger) than $H_{inf}$ (arrow drawn for 3 K).
} 
\efig

However, when the thermal Hall conductivity $\kappa_{xy}$ is large 
(as in CeCoIn$_5$ below 20 K~\cite{onose}), 
a sizeable transverse gradient $-\partial_y T$ appears, which 
drives a charge current $||\bf\hat{y}$ via $\alpha$ (this is just the ordinary thermopower
responding to $-\partial_y T$).  Instead of Eq. \ref{en}, we have~\cite{wang01}
\be
e_N = \rho\alpha_{xy} - S \left[\frac{\sigma_{xy}}{\sigma} + 
\frac{\kappa_{xy}}{\kappa_e}\right],
\label{en2}
\ee
which expresses balancing the 3 off-diagonal currents $\sim\alpha_{xy}$, $\sim\sigma_{xy}$
and $\sim \kappa_{xy}$ by the current $\sigma E_y$.  Thus to obtain
$\alpha_{xy}$ from $e_N$, we should subtract the charge and thermal Hall currents.
(We adopt the sign convention~\cite{wang01,wang06} 
that $e_N$ shares the sign of $\alpha_{xy}$, i.e. 
when $\alpha_{xy}>0$ and dominant, $e_N$ is positive.
Vortex flow in a superconductor gives a positive $e_N$.)

We first describe the field dependence of $S$ and the inferred $\alpha$.  
At low $T$, $S$ displays a very interesting field dependence (Fig. \ref{SH}a). 
Above $\sim$15 K, $S$ is nearly insensitive to $H$ aside from a 
weak cusp-like anomaly at $H$= 0. With decreasing $T$, the anomaly 
deepens and imparts a strong $H$ dependence that extends to $\sim$12 T.  
In the interval 3-7 K, the zero-$H$ cusp is bracketed by a steep $H$-linear 
dependence (at 3 K, $S$ increases 14-fold between $H$ = 0 and 12 T).
Below $T_c$, $S$ rises nearly vertically from zero at $H_{c2}$ (the upper critical field), 
attains a broad maximum (8.1 T at 1 K), and settles to
a plateau value at large $H$.

Using the magnetoresistance $\rho$ vs. $H$ from Ref.~\cite{onose}, we
have converted the $S$-$H$ curves into $\alpha$ vs. $H$ (Fig. \ref{SH}b).
At 17 K, $\alpha$ displays the weak $H^2$ dependence characteristic of
the normal-state thermoelectric conductivity
$\alpha^n$ given by the Mott expression
$\alpha^n = \frac{\pi^2}{3}\frac{k_B^2T}{e}\left[\frac{\partial\sigma}
{\partial\epsilon}\right]_{\mu}$,
where $k_B$ is Boltzmann's constant, $e$ the elemental charge,
$\epsilon$ the energy and $\mu$ the Fermi level. 

However, as $T\rightarrow$3 K, the minima in $\alpha$ deepen to cusps even
sharper than those in $S$.  Below 15 K, the curve of $\alpha$ displays 
an inflexion point at a field $H_{inf}$ where $\partial^2\alpha/\partial H^2 = 0$  
($H_{inf}$ increases from 3.7 T at 3 K to 9.4 T at 9 K).  The inflexion field
reveals 2 factors that pull $\alpha$ in opposite directions.  
As $T$ decreases, $\alpha$ increases if $|H|>H_{inf}$, but decreases for $|H|<H_{inf}$.  
The cusp anomaly is dominant for $|H|<H_{inf}$.
Further, the field scale defining the cusp sharpness ($\sim$100 G at 3 K)
suggests that the field is spoiling the phase coherence of the wave
function at low $T$ (see below).

A notable feature is that the anomaly in $\alpha$ is
much larger than that in $\rho$. For e.g., in the interval
$H= 0\rightarrow H_{inf}$ defining the anomaly, $\alpha$ increases more than 
five-fold at 3 K, whereas the cusp in $\rho$ constitutes only 20$\%$ of the total
resistivity~\cite{onose}.  The unusually large cusps in $S$ reflects
this huge discrepancy.  We return to these features after describing the Nernst results.

\bfig[h]			
\incl[width=9.5cm]{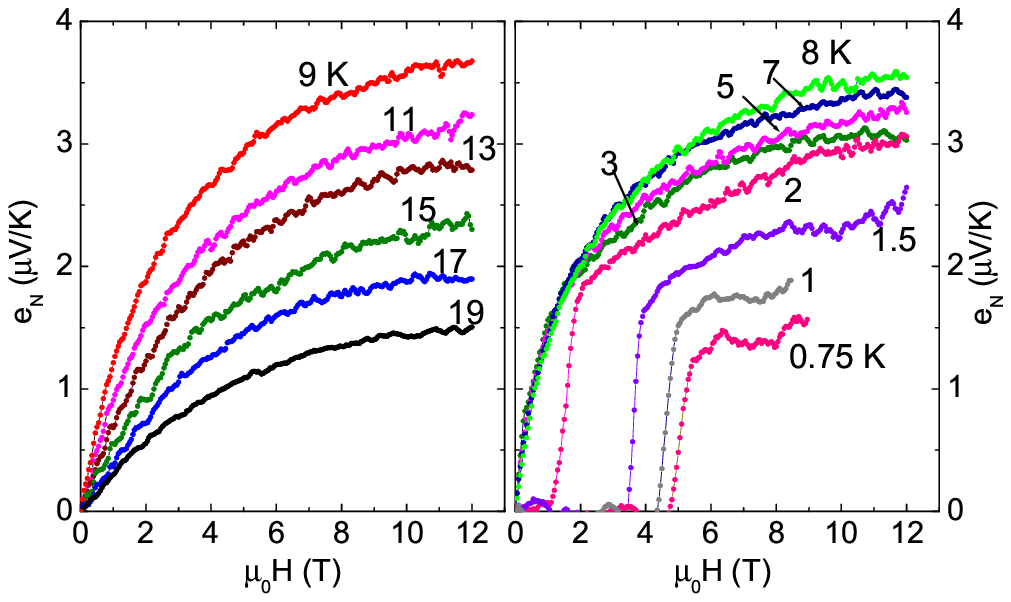}
\caption{\label{eN}  The observed Nernst signal $e_N$ vs. $H$ in CeCoIn$_5$ 
from $T$ = 9 to 19 K (Panel a) and from 0.75 to 8 K (b).  Below $T_c$, $e_N$
undergoes a sharp jump from zero at $H_{c2}$.  Throughout, $e_N$ includes 
a significant contribution from a current driven by $\kappa_{xy}$. The
sign of the Nernst effect is positive.
}
\efig

\bfig[h]			
\incl[width=9.5cm]{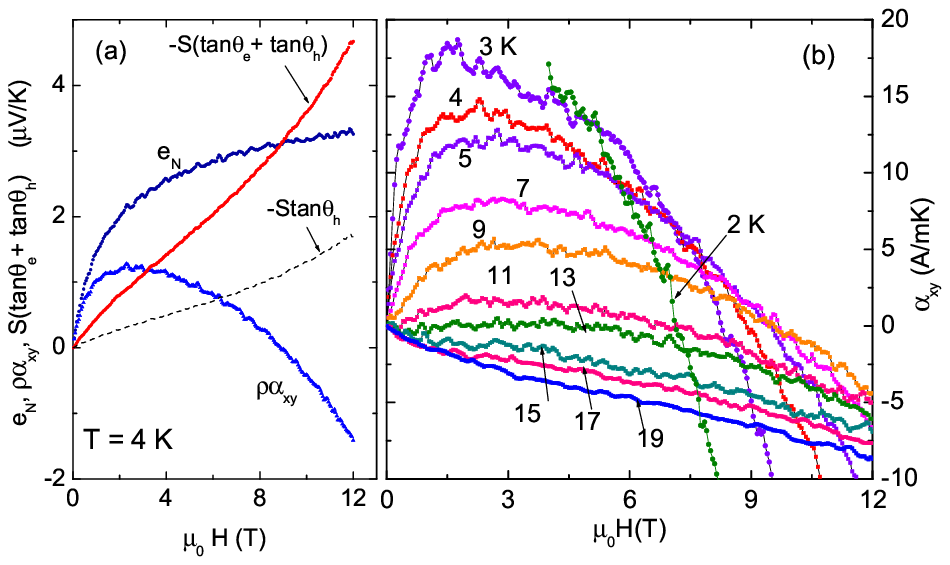}
\caption{\label{axy} (Panel a) Separation of the observed Nernst curve $e_N$ at 4 K into 
the off-diagonal Peltier term $\rho\alpha_{xy}$ and the charge-and-thermal Hall term 
$-S(\frac{\sigma_{xy}}{\sigma} + \frac{\kappa_{xy}}{\kappa})$.  As shown by the dashed
curve, 30-40 $\%$ of the latter derives from $\kappa_{xy}$. $S>0$ (hole-like), while
$\sigma_{xy}<0$ and $\kappa_{xy}<0$ (electron-like).
(Panel b) The curves of $\alpha_{xy}$ vs. $H$ derived from
$e_N$ using Eq. \ref{en2}.  At 17 and 19 K,
$\alpha_{xy}$ is nearly $H$-linear, consistent with a qp
origin.  The plotted quantity is the sum of 2 terms (Eq. \ref{axy2}).
The broadly peaked profile at lower $T$ ($<$15 K) is 
the anomalous term $\alpha_{xy}^s$.  At the lowest $T$, the negative qp 
term $\alpha_{xy}^n$ pulls $\alpha_{xy}$ to large negative values (-100 A/mK at 12 T and 2 K).
}
\efig

In Ref. \cite{bel}, an enhanced Nernst signal was reported below $\sim$25 K, 
but with a sign opposite to that of the vortex-Nernst effect~\cite{xu,wang01,wang06}.  
Our measurements of $e_N$ (Fig. \ref{eN}) are nominally consistent in magnitude with Ref. \cite{bel}.  
However, we find that the sign of $e_N$ is positive (after some correspondence, 
the authors of Ref. \cite{bel} have confirmed that the sign is positive~\cite{izawa2}).  

Using $S$ and the tensors {\boldmath $\sigma$} and {\boldmath $\kappa$}
measured in Ref. \cite{onose}, we now use Eq. \ref{en2} to extract $\alpha_{xy}$.
The separate contributions are shown in Fig. \ref{axy}a at $T$ = 4 K.  
The curve of $e_N$ is the sum 
of $\rho\alpha_{xy}$ and the augmented Hall term 
$-S(\frac{\sigma_{xy}}{\sigma} + \frac{\kappa_{xy}}{\kappa})$.  
We note that, in CeCoIn$_5$, $S>0$ (hole-like), while
$\sigma_{xy}<0$ and $\kappa_{xy}<0$ (electron-like). 
Roughly $\frac13$ of the Hall term derives from the current due to $\kappa_{xy}$ (dashed line).

The derived field profiles of $\alpha_{xy}$ are highly instructive (Fig. \ref{axy}b).
At the high-$T$ end (17 and 19 K), the curves are $H$-linear, consistent with the
off-diagonal thermoelectric response of quasi-particles in moderate fields.
We identify this as the qp background term, atop of which a positive contribution to $\alpha_{xy}$ emerges as $T\rightarrow T_c$.  Referring back to $e_N$ in Fig. \ref{eN}, we may now see that, at temperatures above 15 K, the ``enhanced'' Nernst signal with 
pronounced curvature merely reflects the large charge and thermal Hall currents 
$\sigma_{xy}$ and $\kappa_{xy}$.  These contribute to the Nernst signal, but are not 
intrinsic to the off-diagonal thermoelectric response.

\bfig[h]			
\incl[width=9.5cm]{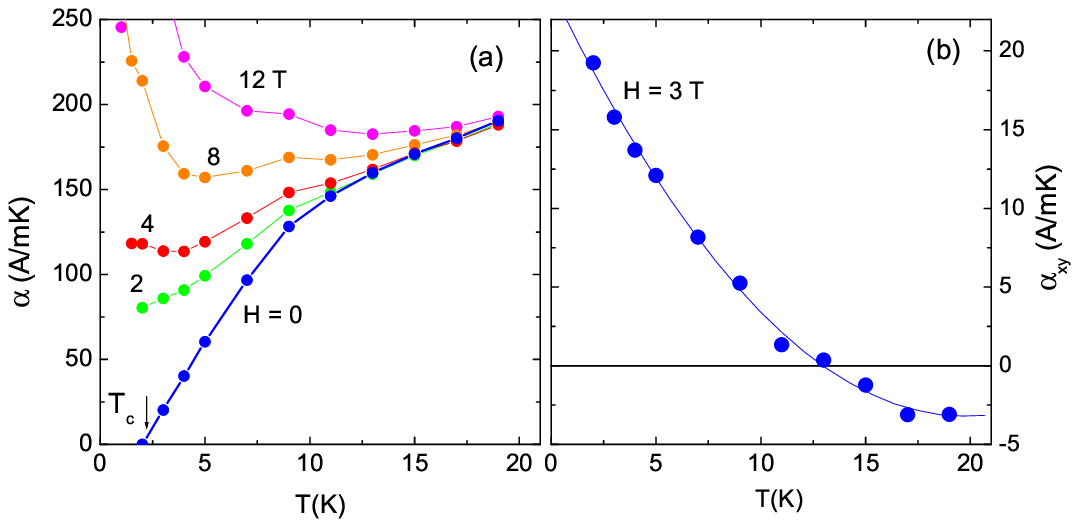}
\caption{\label{aT}  (Panel a) The $T$ dependence of $\alpha$ with $H$ 
fixed at selected values.  At $H$ = 0, the curve of $\alpha$ decreases monotonically
to zero at $T_c$.  A weak $H$ readily suppresses this downward trend.  At 
large $H$ ($>$ 5 T), $\alpha$ is strongly enhanced because of steep
increase of $\ell$.  
Panel (b) shows the $T$ dependence of $\alpha_{xy}$ with $H$ fixed
at 3 T.  The increase in $\alpha_{xy}$ below 15 K mirrors the
decrease in $\alpha$ in weak $H$ in Panel (a).  
Above 15 K, $\alpha_{xy}$ approaches the negative qp term 
$\alpha_{xy}^n$.
}
\efig

The pattern of the curves in Fig. \ref{axy}b suggests that $\alpha_{xy}$ is 
comprised of 2 terms, viz.
\be
\alpha_{xy} = \alpha^n_{xy}+\alpha^s_{xy}.
\label{axy2}
\ee
where $\alpha_{xy}^n$ is the negative qp term and $\alpha_{xy}^s$ the 
positive anomalous term.  Below 15 K, the latter swells rapidly with a characteristic
field profile that peaks at relatively low fields (2-3 T) and then decays slowly
at large $H$.  The $T$ dependence of $\alpha_{xy}$ at a fixed $H$ (3 T) is shown in
Fig. \ref{aT}a.  It is apparent that the anomalous term appears as a positive 
contribution on top of a negative, $H$-linear background.  

In Fig. \ref{axy}b, the curves below 5 K exhibit a steep decrease to large negative
values at high fields.  As reported in Ref. \cite{onose}, 
measurements of $\kappa_{xy}$ and $\sigma_{xy}$ reveal that, below 6 K, the electronic
mean-free-path $\ell$ is sharply enhanced in large $H$ because of
field-suppression of scattering by spin disorder~\cite{onose}.
We identify the high-field changes in $\alpha_{xy}$ with strong enhancement 
of the qp term $\alpha_{xy}^n$ which scales like $\ell^2$ ($\alpha_{xy}^n$ satisfies the Mott expression as $\alpha^n$, with $\sigma$ replaced by $\sigma_{xy}$).
The increase in $\ell$ strongly enhances $\alpha_{xy}^n$, which is 
intrinsically negative, as we noted at 17-19 K.  
At 2 K, the qp contribution is so large that it 
pulls $\alpha_{xy}$ to very large negative values (-100 A/mK at 12 T).  
Hence, at high fields, the total off-diagonal term $\alpha_{xy}$ is dominated by 
the steep growth of the negative qp term, but at low fields, it is 
dominated by the positive anomalous term.  This is just the 2 disparate trends 
separated by $H_{inf}$ in the curves of $\alpha$ (Fig. \ref{SH}b), but now observed
in the off-diagonal channel. 

We now return to the implications of the thermopower. 
Figure \ref{aT}a shows the $T$ dependence of 
$\alpha(T,H)$ at several fixed $H$.  Starting at 20 K, the zero-field
curve $\alpha(T,0)$ initially decreases with a modest slope, but below 15 K,
it accelerates to fall steeply to zero close to $T_c$.  

There are 2 unusual aspects of $\alpha(T,H)$.  First, in 
conventional superconductors, $\alpha$ (and $S$) 
display a step-like decrease to zero at $T_c$, whereas $\alpha$ here
is already fully suppressed just above $T_c$.  By Onsager reciprocity, $\alpha = \tilde{\alpha}/T$, where $\tilde{\alpha}$ defines the heat 
current ${\bf J}_h$ produced by $\bf E$ (with $\nabla T$ = 0), viz.
$J_h = \tilde{\alpha}E$, so that we may regard $J_S = \alpha E$
as the entropy current density generated by $\bf E$ (whence $S$ is the entropy
transported per unit charge). In this view,
the sharp decrease of $\alpha(T,0)$ in the broad 13-K interval above $T_c$
(Fig. \ref{aT}a) implies a dramatic loss of entropy carried by the conduction electrons.  
The loss in carrier entropy is a new feature of the precursor state above $T_c$ in CeCoIn$_5$.

Secondly, as shown in Fig. \ref{SH}b, the zero-field behavior
is very field sensitive.  At 3 K, the cusp is rounded in a very weak field (100 G) 
and nearly completely suppressed at $H_{inf}$ (3.7 T).  A field of 100 G 
is equivalent to a magnetic length $\sqrt{\hbar/eB}\sim$ 2,600 \AA.  It is 
unlikely that field sensitivity on such long lengths can be explained 
semi-classically.  Instead, it is indicative of field decoherence of an
electronic wave function that retains phase coherence over large distances.  
Two examples displaying such weak-field sensitivity are the suppression 
of weak Anderson localization, via dephasing of paths related by 
time-reversal symmetry, and field suppression of phase coherence 
in the fluctuation regime of superconductors.

Within our accuracy, the onset temperature of $\alpha_{xy}^s$
is identical with that of the cusp anomaly in $\alpha$ (Fig. \ref{SH}b).
The actual anomalous contribution detected in the Nernst signal shares the same origin
as the anomalous term in the thermopower; both appear below 15 K rather than
25 K. An important feature, however, is that $\alpha$ is much 
more field sensitive than $\alpha_{xy}^s$, as may be seen
by comparing Figs. \ref{SH}b and \ref{axy}b.  The latter survives to fields larger than 6 T 
whereas the cusp in $\alpha$ is affected by very weak $H$.

\bfig[h]			
\incl[width=7cm]{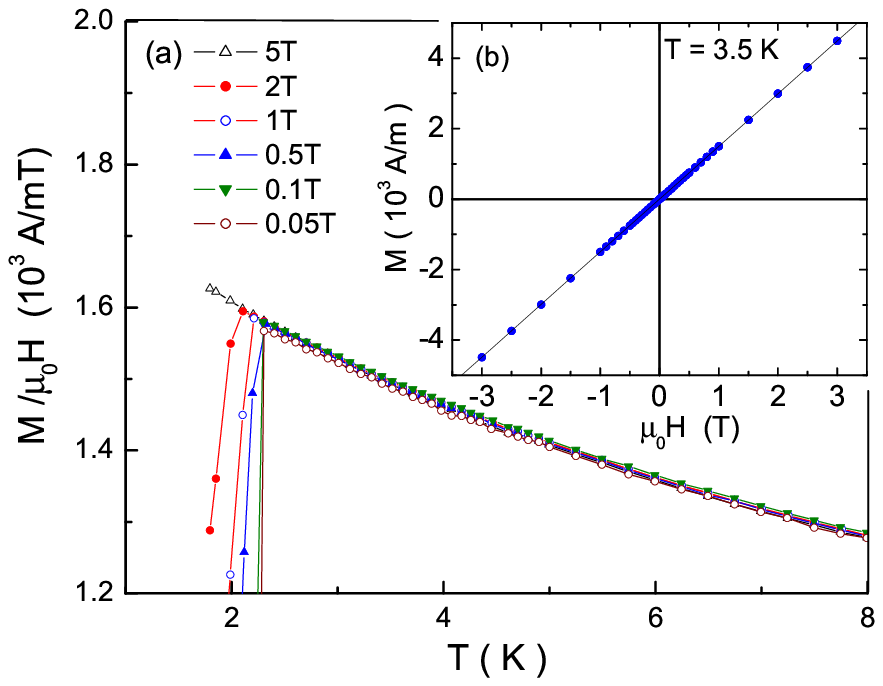}
\caption{\label{MH}
(a) The magnetization $M$ vs. $T$ above $T_c$ in CeCoIn$_5$ 
expressed as a susceptibility $M/\mu_0H$.  Curves measured by SQUID magnetometry 
at fields ${\bf H||c}$ from 0.05 T to 5 T coincide.  
Panel (b) shows the strictly $H$-linear variation of $M$ at 3.5 K.  
The susceptibility $\chi$ equals +1.879$\times 10^{-3}$. Within our resolution,
$M$ is linear in $H$; an $H$-dependent, diamagnetic contribution 
is not detected.  
}
\efig

In analogy with the cuprates~\cite{wang01,wang06}, an appealing 
candidate for the precursor state above $T_c$ 
would seem to be the vortex-liquid scenario in which the Cooper-pair condensate 
lacks phase rigidity on long length scales because of spontaneous
(anti)vortices.  The scenario accounts for the steep fall of $\alpha$ 
as $T\rightarrow T_c$ because regions of the sample in which the condensate retains phase 
coherence have reduced entropy.  In a field, vortices inserted 
by $H$ lead to rapid phase decoherence of these
regions to produce the steep increase observed in $\alpha$.  
Moreover, the flow of vortices in the applied gradient 
generates a large, positive Nernst signal as observed.  
Insofar as the anomalous term $\alpha_{xy}^s$ comes from phase-slip events
arising from the passage of individual vortices, the Nernst signal does rely
on having regions with long-range phase coherence, so it survives to large fields.   
This scenario is compatible with the rapid changes in $\alpha$ in weak $H$, 
as well as the relative robustness of $\alpha_{xy}$ to moderately large fields.

The vortex liquid above $T_c$ should exhibit a sizeable diamagnetic 
signal, as has been confirmed in the case of cuprates~\cite{wang05}.  However, our
measurements of magnetization $M$ have not uncovered such diamagnetism in CeCoIn$_5$
(Fig. \ref{MH}a).  Above $T_c$, the magnetization is dominated by a $T$-dependent, 
paramagnetic susceptibilty that is large ($\chi\sim 10^{-3}$)~\cite{mag}.  
Within our resolution, $M$ is observed to be strictly linear 
in $H$ from 0 to 5 T at 3.5 K, with no trace of a diagmagnetic fluctuation
contribution (Fig. \ref{MH}b). Moreover, below $T_c$, the 
upper critical field $H_{c2}$ is sharply defined and given by the mean-field form
$H_{c2}\sim (T_c-T)$, very unlike the situation in cuprates.  Clarification of 
the magnetization may require experiments that can resolve a fluctuating
$M$ at the level of a 1-10 A/m~\cite{wang06}.

The thermoelectric current response in a field has revealed
several unusual characteristics of the precursor state in CeCoIn$_5$ which
appear below 15 K. In zero $H$, the striking decrease of $S$ 
and $\alpha$ towards $\sim$0 near $T_c$ implies a sharp loss of the 
entropy current carried by quasiparticles.  The striking field sensitivity of $\alpha$ 
implies that the zero-entropy feature occurs in regions with long-range 
phase coherence.  Concurrently, the off-diagonal current $\alpha_{yx}(-\nabla T)$
gains an anomalous positive contribution, which is readily distinguised
from the negative qp contribution.  In contrast to these very prominent anomalies
in the thermoelectric current, the modification to the resistivity
is quite modest~\cite{onose}.  The contribution to the magnetization is currently below our
resolution.

We acknowledge valuable disussions with K. Behnia, Y. Matsuda, 
C. Broholm and L. Taillefer.  Y. O. thanks the Nishina Memorial 
Foundation for partial support.  Research at Princeton University 
and the Brookhaven National Laboratory was supported by the U.S. 
National Science Foundation (DMR 0213706) and the 
U.S. Department of Energy (DE-Ac02-98CH10886), respectively.

\end{document}